\documentclass[epjST]{svjour}

\usepackage[numbers]{natbib}

\usepackage{graphicx}        % standard LaTeX graphics tool  

\def\newblock{\hskip .11em plus .33em minus .07em}

\newcommand{\apj}{\mbox{ApJ}}

\newcommand{\mnras}{\mbox{MNRAS}}

\newcommand{\nat}{\mbox{Nature}}

\newcommand{\pasp}{\mbox{PASP}}
\newcommand{\pasj}{\mbox{PASJ}}
\newcommand{\na}{\mbox{New Astronomy}}
\newcommand{\nar}{\mbox{New Astronomy Reviews}}

\newcommand\bain{\mbox{Bull.~Astron.~Inst.~Netherlands}}% Bul. Astr. Inst. of the Netherlands 
\newcommand\zap{\mbox{ZAp}}% Zeitschrift fuer Astrophysik 

\def\apgt{\ {\raise-.5ex\hbox{$\buildrel>\over\sim$}}\ }
\def\aplt{\ {\raise-.5ex\hbox{$\buildrel<\over\sim$}}\ }
\def\lt{\ {\raise-.5ex\hbox{$\buildrel>$}}\ }
\def\gt{\ {\raise-.5ex\hbox{$\buildrel<$}}\ }

\begin{document}

\title{A pilgrimage to gravity on GPUs}
\author{Jeroen B\'edorf\inst{1}\fnmsep\thanks{\email{bedorf@strw.leidenuniv.nl}} \and 
Simon Portegies Zwart\inst{1}}
\institute{$^1$Leiden Observatory, Leiden University, P.O. Box 9513, 2300 RA Leiden, 
The Netherlands }

\abstract{
In this short review we present the developments over the last 5 decades that have led to 
the use of Graphics Processing Units (GPUs) for astrophysical simulations. Since the
introduction of NVIDIA's Compute Unified Device Architecture (CUDA) in 2007 the 
GPU has become a valuable tool for $N$-body simulations and is so popular 
these days that almost all papers about high precision $N$-body simulations 
use methods that are accelerated by GPUs. With the GPU hardware becoming more
advanced and being used for more advanced algorithms like gravitational tree-codes
we see a bright future for GPU like hardware in computational astrophysics.
%TODO: Fix the reference format section...weird stuff
}

\maketitle

\section{Introduction}

In this review we focus on the hardware and software developments since the 1960s that 
led to the successful application of Graphics Processing Units (GPUs) 
for astronomical simulations. The field of $N$-body simulations is 
broad, so we will focus on direct $N$-body and hierarchical
tree-code simulations since in these two branches of $N$-body simulations
the GPU is most widely used. We will not cover cosmological simulations
despite this being one of the computationally most demanding branches
of $N$-body simulations, however the GPU usage is negligible. 

There are many reviews about $N$-body simulations which all
cover a specific branch or topic, some of the most recent reviews
are those by Dehnen and Read~\cite{2011EPJP..126...55D} with a focus on
used methods and algorithms as well as the 
work of Yokota and Barba~\cite{2011arXiv1108.5815Y} which
especially focus on Fast Multipole Methods and their implementation on GPUs. 

\begin{figure*}
\includegraphics[width=0.75\columnwidth,angle=-90]{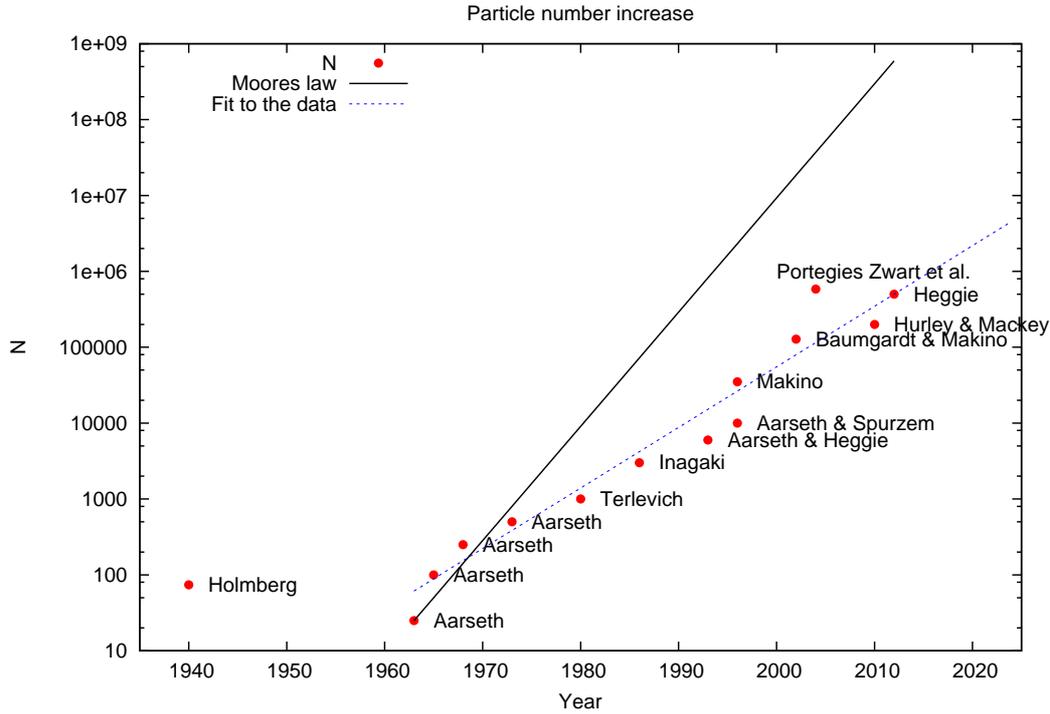}
 \caption{Number of particles used in collisional simulations 
over the last 4 decades. The solid line shows Moore's law~\cite{Moore}, the circles 
publications and the dashed-line a fit through the data points.(Adapted from~\cite{2010NewAR..54..163H, 2003gmbp.book.....H})}
 \label{fig:NIncrease}        
\end{figure*}

In this review we follow a chronological approach divided
into two parallel tracks: the collisional direct $N$-body methods
and the collisionless tree-code methods. For the direct simulations 
we partially follow the papers mentioned by D. Heggie and 
P. Hut~\cite{2010NewAR..54..163H, 2003gmbp.book.....H}
as being noteworthy simulations since the 1960s. These papers and the number of bodies
used in those simulations are presented in Fig.~\ref{fig:NIncrease}
(adapted from~\cite{2010NewAR..54..163H, 2003gmbp.book.....H});
in the figure we show the number of bodies used and Moore's law which 
is a rough indication of the speed increase of computer chips~\cite{Moore}.

\begin{figure*}
\includegraphics[width=0.75\columnwidth, angle=-90]{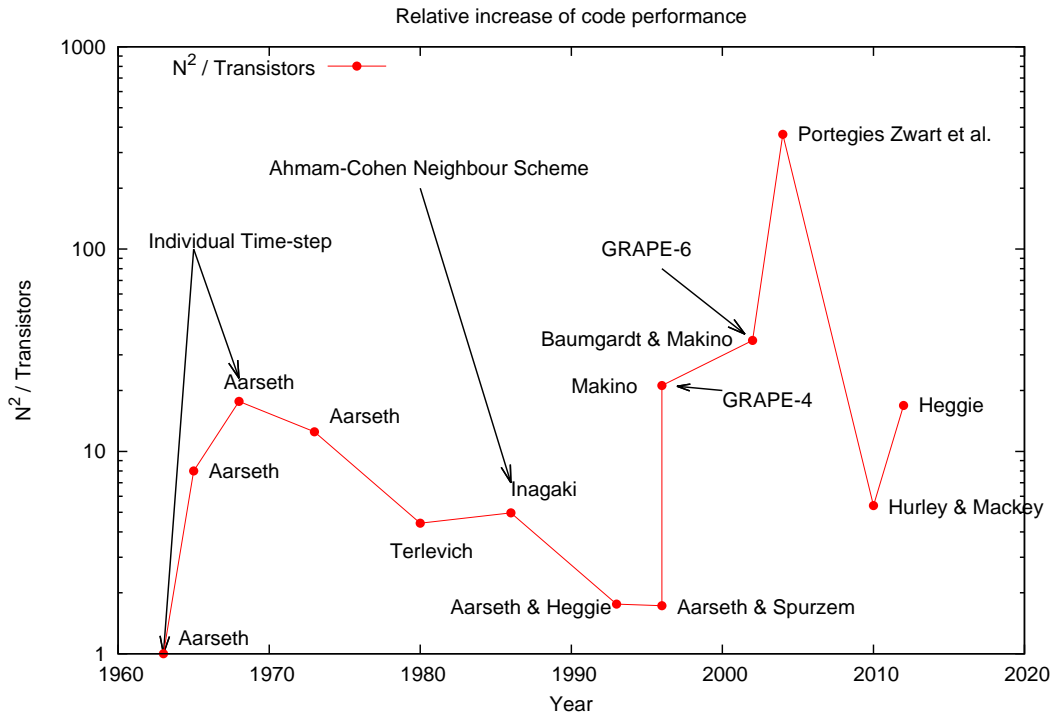}
 \caption{The number of theoretical operations, $N^2$, divided
          by the number of transistors determined using Moore's law 
          where the start year is set and normalized to 1963 }
 \label{fig:NScale}        
\end{figure*}

Since direct $N$-body methods scale as ${\cal O}(N^2)$ it is understandable that 
the number of bodies used do not follow Moore's law in Fig.~\ref{fig:NIncrease}.
However, in Fig.~\ref{fig:NScale} we show that the increase in the number
of bodies is faster than would be explainable by increase in computer speed
alone. We show the theoretical number of operations, which
is $N^2$ in the naive situation. And the number of transistors which is
an indication of the speed of the computer, whereby we set the start year to
1963. If the increase in $N$ was solely based on the increase in computer
speed the line would be horizontal and equal to 1. Everything above 1 
indicates that the improvement comes from software or 
hardware of which the speed doubles every 18 months according to 
Moore's law. In the figure we tried to indicate (with arrows) what 
the major reasons for improvements were.

\section{The very beginning}

\begin{figure*}
\includegraphics[angle=90,width=\columnwidth]{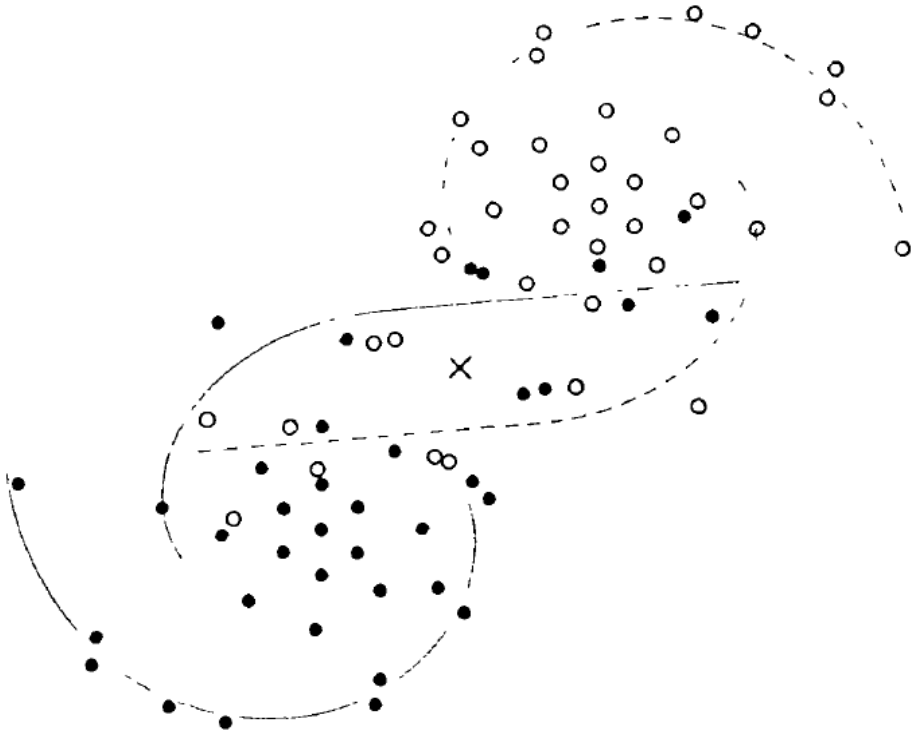}
 \caption{Spiral arms formed in the experiments by Holmberg in 1941. 
    Image taken from~\cite{1941ApJ....94..385H}}
 \label{fig:Holmberg}        
\end{figure*}

Before the first computer simulations the Swedish astronomer Erik
Holmberg \cite{1941ApJ....94..385H} published simulations of two
interacting galaxies which were conducted using light bulbs. In his
experiment each galaxy was represented by 37 light bulbs. Holmberg
then measured the brightness of the light bulbs, which falls off with
$1\over r^{2}$, to compute the gravitational forces and let the
galaxies evolve.  In Fig.~\ref{fig:Holmberg} we show one of his 
results where spiral arms develop, because of the interactions 
between two galaxies.
This experiment was specifically tailored for one
problem, namely gravitational interaction, which made it difficult to
repeat using other more general hardware available at the time. So, even
though it took a lot of manual labour, it would take almost
20 years before digital computers were powerful enough to perform
simulations of comparable size and speed. This is the advantage of
tailoring the hardware to the specific problem requirements. 50 years
later we see the same advantage with the introduction of the special
purpose GRAPE hardware (see Section~\ref{Sect:2000-today}).

\section{1960 - 1986: The Era Of Digital Computers}
\label{Sect:1960-1986}

The introduction of general purpose digital computers in the 1960s
made it easier to buy and use a computer to perform simulation of
$N$-body systems. The digital computers were based on transistors
instead of vacuum tubes which made them cheaper to produce and
maintain. The first computer simulations of astrophysical $N$-body
systems where performed by von Hoerner in
1960~\cite{1960ZA.....50..184V}, Aarseth in 1963~\cite{1963MNRAS.126..223A} and van Albada 
in 1968~\cite{1968BAN....19..479V}.  The number of bodies involved in these
simulations was still relatively small and comparable to the
experiment of Holmberg ($N=10$ to $100$).

During the first decades of the digital computer there were two ways
to increase the number of bodies. One method was to buy
a faster computer
which allowed you to keep using the same software but increase the
number of used particles. This was an efficient method in the sense
that the speed of the computer doubled roughly every 18 months, following
Moore's Law~\cite{Moore}\footnote{Technically Moore does not describe
the speed of the computer, but the number of transistors. In practice the
  speed of a computer chip is roughly related to the number of
  transistors.}.  Another method to increase $N$ was by improving the
software either by code optimizations or by algorithmic changes. In
direct $N$-body integrations the number of required interactions
scales as $N^2$ so any improvement to reduce the number of operations
is very welcome.

In 1963 Aarseth\cite{1963MNRAS.126..223A} introduced the individual
time-step scheme.  To simulate an $N$-body system the orbits of
particles have to be followed exactly. However, particles isolated in
space do not have sudden changes in their orbit and therefore can take
longer time-steps than particles in the core of the cluster or which
are part of a binary. Particles forming a binary change their positions
quickly and therefore require many more time-steps to be
tracked accurately. This is the basic idea behind the individual
time-step scheme, each particle is assigned a simulation time when it
is required to update and recompute the gravitational force.  When
only a few particles take small time-steps
then for most of the particles the gravitational forces do not have to
be computed. Thereby going from $N^2$ operations in a shared
time-step scheme to $N\cdot N_{\rm active}$ operations where $N_{\rm
  active}$ is the number of particles that have to be updated.  If
$N_{\rm active}$ is sufficiently small then the overhead of keeping
track of the required time-steps is negligible compared to the gain in
speed by not having to compute the gravitational forces for all bodies
in the system.

The Ahmam-Cohen Neighbour Scheme (ACS) introduced in 1973~\cite{1973JCoPh..12..389A}
takes another approach to reduce the number of required
computations. In ACS the gravitational force computation is split into
two parts. In the first part the force between a particle and its
nearest neighbours, $N_{nn}$, (hence the name) is computed in a way
similar to the direct $N$-body scheme with many small time-steps,
because of the fast changing dynamical nearby neighbourhood. In the
second part the force from the particles that are further away is
updated less frequently, since the changes to that part of the
gravitational force are less significant.  When $N_{nn} \ll N$ the
number of total interactions is reduced dramatically and thereby the
total wall-clock time required for the simulation is reduced.

With the introduction of the digital computer also came the
introduction of parallel computing. You can distinguish fine grained
and coarse grained parallelisation. The former focuses on tasks that
require many communication steps whereas the latter splits the
computational domain and distributes it among different processors.
These processors can be in the same machine or connected by a network.
When connected by a network the communication is slower and therefore 
only beneficial if the amount of communication is
minimal. In the early years of computing the focus was on fine grained
parallelism using vector instructions. These instructions helped to increase the
number of bodies in the simulations, but still $N$ increased much
slower than the theoretical speed of the processors. 
This is because of the $N^2$ scaling of direct $N$-body
algorithms. Some of the noteworthy publications were the simulation of
open clusters containing 1000 stars by Terlevich in 1980~\cite{1980IAUS...85..165T}
and the simulation of globular clusters using up to 3000 particles by Inagaki in
1986~\cite{1986PASJ...38..853I} using the (at the time) commonly used
{\tt NBODY5} code.

% (during that time
% computers were still single-core so no on-chip parallelism existed except for vector
% instructions). 

\section{1986 - 2000 : Advances in software}
\label{Sect:1986-2000}

In 1986 Barnes \& Hut ~\cite{1986Natur.324..446B} introduced a
collisionless approximation scheme based on a hierarchical data
structure.  With the introduction of the BH Tree-code we see two
parallel tracks in $N$-body simulations: the first, using high
precision direct $N$-body methods and the second using approximation
methods thereby allowing for larger particle simulations.  For more 
information about the
collisionless methods see Section~\ref{Sect:Collisionless}.

The individual time-step method reduced the total number of executed
gravitational force computations, since particles are only updated when
required. However, if you would use the individual time-step method in
a predictor-corrector integration scheme\footnote{In
a predictor-corrector scheme the positions are updated in multiple steps.
First you predict the new positions using the original computed gravitational forces,
then you compute the new gravitational forces using these positions and 
then you apply a correction to the predicted positions.} you would have to predict all
$N$ particles to the new time while only computing the gravitational force for one
particle. This prediction step results in a large overhead and the possibilities to
parallelise the algorithm are limited. Since the gravitational force is computed
for only one particle. In a shared time-step method there 
are $N$ particles for which the force is computed which can then be divided over
multiple processors. A solution came in the form of the block
time-step method in which particles with similar time-steps are grouped together.
These groups are then updated using a time-step that
was suitable for each particle in the group. Since multiple particles
are updated at the same time, the number of prediction steps is reduced and
the amount of parallel work is increased~\cite{1986LNP...267..156M}.
This is an example of $i$-particle parallelisation in which all
$j$-particles are copied to all nodes and each node works on a subset
of the $i$-particles. The $i$-particles are the sinks
and the $j$-particles are the sources for the gravitational forces. In
hindsight, it might have been more efficient to use $j$-particle
parallelisation in which each processing node would get a part of the
total particle set. The $i$-particles that have to be updated during a 
time-step are then broadcast to each node. The nodes then compute 
the gravitational force on those $i$-particles using their subset of 
$j$-particles and finally in a reduction step these partial forces are
combined. With the introduction of the
GRAPE hardware a few years later (see below), it turned out that this $i$-particle
parallelisation was ideal for special purpose hardware.

\begin{figure*}
 \includegraphics[width=\columnwidth]{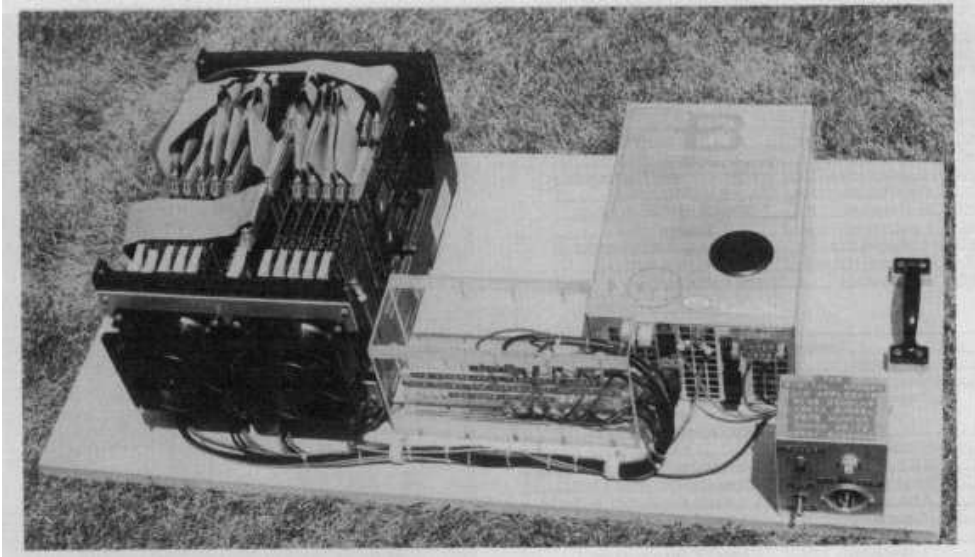}
 \caption{The digital Orrery. Image taken from~\cite{Applegate:1985:DO:4135.4141}}
 \label{fig:DigitalOrrery}        
\end{figure*}

The main focus in the development of $N$-body codes was on increasing
$N$ in order to get increasingly detailed simulations, although some
research groups focused on specialized problems like planetary
stability. The group of Gerry Sussman developed a special machine just
for integrating the solar system, The {\tt Digital
  Orrery}~\cite{Applegate:1985:DO:4135.4141}.  The machine consisted
of a set of specially developed computer chips placed on extension
boards which were connected using a special ring network. A photo of
the machine is shown in Fig.~\ref{fig:DigitalOrrery}. With this machine
the developers were able to find previously unknown chaotic motions in the orbit of
Pluto, caused by resonance with Neptune~\cite{1988Sci...241..433S}.

In 1993 Aarseth \& Heggie~\cite{1993ASPC...48..701A} published the
results of a 6000 body simulation containing primordial binaries and
unequal mass particles. With this simulation it was possible to
improve on previous results where only equal mass particles were used.
The differences in the unequal mass and equal mass simulations were
small, but too large to be ignored, which shows the critical importance of
binaries and initial mass functions even though they are
computationally expensive.

Spurzem and Aarseth~\cite{1996MNRAS.282...19S} performed a simulation
of a star cluster with $10^4$ particles in 1996. The simulation
was executed on a CRAY machine. This is one of the last simulations
in our review that was executed without any special purpose
hardware. Since in the same year, Makino et al.~\cite{1996ApJ...471..796M}
presented their work which used three times more particles and was execute
on GRAPE hardware.

\section{2000 - 2006: The Era Of The GRAPE}
\label{Sect:2000-today}

The introduction of the special purpose GRAvity PipE (GRAPE) hardware
caused a breakthrough in direct-summation $N$-body simulations
\cite{1998sssp.book.....M}. The GRAPE chips have the gravitational
force calculations implemented in hardware which results in a speed-up
of two orders of magnitude compared to the standard software
implementations.  The GRAPE chips were introduced in the early 1990s
\cite{1991PASJ...43..841F}, but it would take a few years and
development cycles before they were widely accepted and being used in production
simulations. The GRAPE chips are placed on a PCI-expansion card that can be
installed in any general purpose (desktop) computer. The GRAPE came
with a set of software libraries that made it relatively easy to add
GRAPE support to existing simulation software like
NBODY4~\cite{1999PASP..111.1333A} and
Starlab~\cite{2001MNRAS.321..199P}.  
The block time-stepping scheme introduced a few years earlier
turned out to
be ideal for this hardware and when multiple GRAPE chips were used one
could combine this in $i-j$-particle parallelisation.

In the early 90s the large computational cost of direct $N$-body
simulations had caused researchers to start using collisionless codes
like the Barnes-Hut tree-code (see Section~\ref{Sect:Collisionless}) in
order to do large $N$ simulations.  The introduction of the GRAPE
combined with the availability of ready-to-use software
caused the opposite effect since suddenly it was possible to do
collisional simulations at the same speed as collisionless
simulations.
The last generation of the fixed function GRAPE hardware
was the GRAPE-6 chip. These were the most commonly used GRAPE chips
and when placed on a GRAPE-6Af extension board they had a peak
performance of $\sim$131 GFLOPs and enough memory to store 128k
particles.

The GRAPE is designed to offer large amounts of fine grained
parallelism since the on-chip communication is fast and
specifically designed for the gravity computations.  Supercomputers on
the other hand are designed for coarse grained parallelisation thereby
offering a large amount of computational cores connected using fast
networks. But the communication times are still orders of magnitude
slower than on-chip communication networks. This means that
supercomputers are rarely used for direct $N$-body simulations and are
much more suitable for collisionless simulations
(Section~\ref{Sect:Collisionless}).  It would require hundreds of
normal processor cores to reach the same performance as the GRAPE
offers on one extension board and that is even without taking into
account the required communication time.  If this is taken into
account then the execution time on supercomputers is unrealistically
high for high precision (e.g. many small time-steps with few active
particles) direct $N$-body simulations

In the 2000s it was clear that parallelisation had become one of the
requirements to be able to continue increasing the number of
particles, because hardware manufacturers shifted their focus from
increasing the clock-speed to increasing the number of CPU cores and
the introduction of special vector instructions\footnote{Like
  Streaming SIMD Extensions (SSE) and Advanced Vector Extensions
  (AVX)}. 
The clock speed came near the physical limit of the silicon 
and CPUs used so much energy that the produced heat became 
a serious problem.
For direct $N$-body simulations with
individual or block time-steps often a combination of fine grained and
coarse grained parallelisation is used (depending on, for example, the
number of particles that is integrated). When the number of particles
that have to take a gravity step is small then it is more
efficient to not use the external network, but rather let all the work
be handled by one machine. On the other hand if the number of particles 
taking a gravity step is large it could be more efficient to distribute
the work over multiple machines. \\

The introduction of Streaming SIMD Extensions (SSE) vector
instructions in modern day processors promised to give a performance
boost for optimized code. However this optimization step required deep
technical knowledge of the processor architecture. With the
introduction of the Phantom GRAPE library by Keigo
Nitadori~\cite{2006NewA...12..169N} it became possible to benefit from
these instructions without having to write the code yourself.  As the
name suggests the library is compatible with software written for
GRAPE hardware, but instead executes the code on the host processor
using the special vector instructions for increased
performance. Recently this is extended with the new Advanced Vector
eXtensions (AVX) which allows for even higher performance on the
latest generation of CPUs~\cite{2012NewA...17...82T, 2012arXiv1203.4037T}. \\

One of the limitations of the GRAPE is its fixed function pipeline and
because of this it can not be used for anything other than gravity
computations. For example in the Ahmam-Cohen neighbour Scheme the
force computation is split into a near and a far force. The GRAPE
cards are suitable to speed-up the computation of the far force, but
the near force has to be computed on the host since the GRAPE has no
facility to compute the force using only a certain number of
neighbours.  An alternative like Field Programmable Gate
Arrays (FPGAs) allows for more flexibility while still offering high
performance at low energy cost, since the hardware can be programmed
to match the required computations.  The programming is complex, but
the benefits can be high since the required power is usually much
less than a general purpose CPU cluster offering similar performance.
An example of FPGA cards are the MPRACE
cards~\cite{2007JPhCS..78a2071S}, which are designed to speed-up the
computation of the neighbour forces and thereby eliminating the need
to compute the near force on the host computer which would become a
bottleneck if only GRAPE cards would be used.

With the increasing availability of the GRAPE hardware at different 
institutes came the possibility to combine multiple GRAPE clusters 
for large simulations. A prime example of this is the work by 
Harfst et al.~\cite{2007NewA...12..357H} who used two parallel supercomputers
which were equipped with GRAPE-6A cards. They showed that for direct 
$N$-body simulations it was possible to reach a parallel efficiency
of more than 60\% and reached over 3TFLOPs of computational speed
when integrating $2\times10^6$ particles. 
Though the number of GRAPE devices was increasing it was still
only a very small fraction of the number of ``normal'' PCs that was available. 
In order to use those machines efficiently one had to combine them 
and run the code in parallel. An example of this is the parallelisation
of the $N$-body integrator in the {\tt Starlab} package~\cite{2008NewA...13..285P}.
This work showed that it was possible to run parallel $N$-body simulations
over multiple computers, although it was difficult to get good enough 
scaling in order to compete with the GRAPE hardware.  This was also 
observed in earlier work by Gualandris et al.~\cite{Gualandris2007159}
who developed different parallel schemes for $N$-body simulations 
thereby observing that the communication time would become a 
bottleneck for simulations of galaxy size systems.
Another approach is the work by Dorband et al.~\cite{2003JCoPh.185..484D}
in which they implemented
a parallel scheme that uses non-blocking communication. They called this
a systolic algorithm, since the data rhythmically passes through a network
of processors. Using this method they were able to simulate $10^6$ 
particles using direct $N$-body methods.

\section{2006 - Today: The Era Of Commercial High Performance Processing Units}
\label{Sect:2006-today}

\begin{figure}
  \includegraphics[width=\columnwidth]{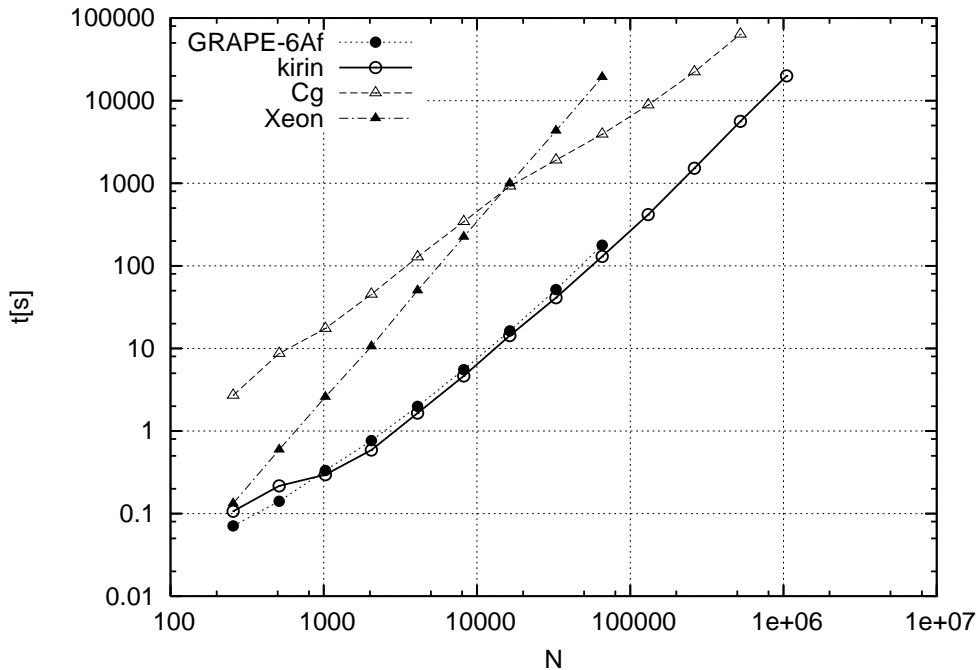}
  \caption{Performance comparison of $N$-body implementations.
    The CUDA GPU implementation  {\tt kirin} is
  represented by the solid line (open circles). The GRAPE is
  represented as the dotted line (bullets).
  The Cg GPU implementation is
  represented as the dashed line (open triangles). The dashed-dotted
  line (closed triangles) represent the results on the host computer.
  Figure taken from ~\cite{2008NewA...13..103B}.}
  \label{fig:GPUBellemanetal}
\end{figure}

With the introduction of programmable Graphics Processing Units (GPU) in
2001 (NVIDIA's Geforce 3) it became possible to program high
performance chips using software. However, it would take another 7
years before GPUs were powerful enough to be a viable alternative to
the fixed function GRAPE that dominated the $N$-body simulation field
over the previous decade.  The GPU was originally designed to improve the
rendering speed of computer games. However, over the years these cards
became progressively faster and, more importantly, they became
programmable.  At first one had to use programming languages which
were specially designed for the creation of visual effects (e.g. {\tt
  Cg} and {\tt OpenGL}).  The first use of the GPU for $N$-body
simulations was by Nyland et al. in 2004~\cite{Nyland04} who used {\tt
  Cg}. Their implementation was mostly a proof-of-concept and lacked
advanced time-stepping and higher order integrations which made it
unsuitable for production quality $N$-body simulations. Programming GPUs
became somewhat easier with the introduction of the
BrookGPU~\cite{Buck:2004:BGS:1186562.1015800} programming language.
This language is designed to be a high level programming language and
compiles to the {\tt Cg} language.  In 2006 Elsen et
al. \cite{Elsen:2006:NSG:1188455.1188649} presented an $N$-body
implementation using the BrookGPU language. Around the same time
Portegies Zwart et al. published an implementation of a higher order
$N$-body integration code with block time-steps written in Cg
\cite{2007NewA...12..641P}. Although these publications showed the
applicability and power of the GPU, the actual programming was still a
complicated endeavor.  This changed with the introduction of the
Compute Unified Device Architecture (CUDA) programming language by
NVIDIA in early 2007.  The language and compatible GPUs were specifically
designed to let the power of the GPU be harvested in areas other than
computer graphics.  Shortly after the public release of CUDA
efficient implementations of the $N$-body problem were
presented \cite{2007astro.ph..3100H, 2008NewA...13..103B,
  Nyland_nbody}. Belleman et al.~\cite{2008NewA...13..103B} showed
that it was possible to use the GPU with CUDA for high order $N$-body
methods and block-time steps with an efficiency comparable to the, 
till then, unbeaten GRAPE hardware (see Fig~\ref{fig:GPUBellemanetal}).

With the GRAPE being around for over 15 years most of the production
quality astrophysical $N$-body simulation codes were using the GRAPE
when NVIDIA released CUDA. It was therefore relatively easy to shift
from the GRAPE to the GPU with the introduction of the GRAPE-compatible
GPU library {\tt Sapporo} \cite{Gaburov2009630}. This
library uses double-single precision\footnote{This technique gives
  precision up to the 14th significant bit while using single
  precision arithmetic.} and on-device execution of the prediction
step, just like the GRAPE-6Af hardware.

The GPU chips are produced in much higher volumes than the GRAPE chips
which makes the GPU more cost efficient to produce and cheaper to
buy.  This, combined with more on-board memory, higher computational
speed and the option to reprogram them to your specific needs, is the
reason that nowadays more GPUs than GRAPEs are used for $N$-body
simulations. Even though the GRAPE, because of its dedicated design,
requires less power than the GPU.

The GRAPE-DR (Greatly Reduced Array of Processor Elements with Data
Reduction) \cite{Makino:2007:GMC:1362622.1362647} is different from the earlier
generation GRAPE chips since it does not have the gravitational
computations programmed in hardware, but rather consists of a set of
programmable processors. This design is similar to how the GPU is
built up, but uses less power since it does not have the overhead of
graphic tasks and visual output that GPUs have. At the time of its
release in 2007 the GRAPE-DR was about two times faster for direct 
$N$-body simulations than the GPU.

In Nitadori \& Aarseth 2012 (private communication) the authors
describe their optimisations to the 
NBODY6 and NBODY7~\cite{2012arXiv1202.4688A} simulation codes to
 make use of the GPU.
 They have tested which parts
of the code are most suitable to be executed on the GPU and came to
the conclusion that it was most efficient to execute the so-called
`regular force' on the GPU. This step involves around 99 percent of the
number of particles. On the other hand the local force is executed on
the host using vector instructions, since using the GPU for this step
resulted in a communication overhead which is too large \footnote{ This is
  similar to how the MPRACE project did the division between the GRAPE
  and MPRACE cards}. In this division the bulk of the work is executed on the GPU and the
part of the algorithm that requires high precision, complex operations
or irregular memory operations is executed on the host machine
possibly with the use of special vector instructions. This is a trend
we see in many fields where the GPU is used. Although
some authors overcome the communication overhead by implementing more 
methods on the GPU besides the force computation~\cite{phiGPU_threecontinents}.

In 2009 the Khronos group\footnote{The Khronos Group is a group of
  companies that develops open standards for accelerating graphics,
  media and parallel computations on a variety of platforms} introduced
the OpenCL programming language. The language is designed to create
parallel applications similar to the way CUDA is used for GPUs, with
the difference that programs written in OpenCL also work on systems
with only host CPUs. The idea behind this is that the developer has
only to write and maintain one software program. In reality, however,
the developer will have to write code that is optimized for one
platform (GPU or CPU) in order to get the highest performance out of
that platform. That CUDA was released a couple of years
earlier, has more advanced features and has more supported libraries than
OpenCL are the reasons CUDA is currently more commonly used than
OpenCL. 
However, there is no reason this cannot change in the future with
updated libraries that offer OpenCL support 
(e.g.  {\tt Sapporo2}, B\'edorf et al. in preparation).

The two most recent large $N$ simulations have been performed
by Hurley and Mackey~\cite{2010MNRAS.408.2353H} ($N=10^5$) and 
Heggie ($N=5\times10^5$)(private communication) who both used NBODY6 
in combination with one or more GPUs.

Simulations that are used to determine planetary stability usually
involve only a few particles. This severely limits the amount of 
parallelism and therefore a different approach has to be taken
than in large $N$ simulations. An example of a method that takes
a different approach is {\tt
  Swarm-NG}\footnote{http://www.astro.ufl.edu/{$\sim$}eford/code/swarm/}.
This is a software package for integrating ensembles of few-body
planetary systems in parallel with a GPU. {\tt Swarm-NG} is
specifically designed for low $N$ systems. Instead of breaking up one
problem into parallel tasks, {\tt Swarm}{\tt -NG} integrates thousands
of few-body systems in parallel. This makes it
especially suited for Monte Carlo-type simulations where the same
problem is run multiple times with varying initial conditions.

\section{Collisionless}
\label{Sect:Collisionless}

\begin{figure*}
 \includegraphics[width=0.75\columnwidth]{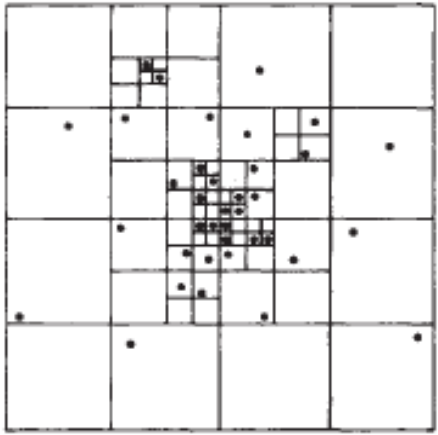}
 \caption{Particles grouped together in boxes in the tree-code algorithm.
          Image taken from~\cite{1986Natur.324..446B}}
 \label{fig:barneshut}        
\end{figure*}

In 1986 Barnes \& Hut introduced their collisionless approximation
scheme based on a hierarchical data structure, which became known as the Barnes-Hut
tree-code~\cite{1986Natur.324..446B}.  In this hierarchical data
structure (tree) the particles are grouped together in boxes (see for
an example Fig.~\ref{fig:barneshut}). These boxes get the combined
properties of the underlying particle distribution, like center of
mass and total mass. To compute the gravitational force on a
particle one does not compute the force between that particle
and all other particles in the system, but rather between the particle
and a selection of particles and boxes. This selection is determined
by traversing the tree-structure and per box deciding if the particle
is distant enough or whether the box lies so close that we should use the
particles that belong to the box. This decision is made by a
`Multipole Acceptance Criterion' which, in combination with the free
parameter $\theta$, is used to get the required precision. In this way one
can either get high precision at high computational costs, by using
more particles than boxes, or the other way around, lower precision by
using more boxes instead of particles. The resulting code that
implements this algorithm generally achieves a scaling of ${\cal O}(N
\log N)$ instead of the ${\cal O}(N^2)$ of direct $N$-body codes.

For large collisionless simulations ($N>10^7$), supercomputers are
required for the available computational resources and for 
the amount of available memory.
Collisionless simulations usually scale as ${\cal O}(N \log N)$ in
tree methods or even as ${\cal O}(N)$ with the Fast Multipole Moment 
(FMM) methods which allows for simulations with much higher
particle numbers~\cite{2002JCoPh.179...27D,2011arXiv1108.5815Y, 2011CoPhC.182.1272Y}.
This requires memory to store the particles, but
also computational resources to solve the gravitational equations. The
gravity equations can usually be solved using specialized hardware,
but those do not have large enough memory buffers to store all
particles.\\

With the introduction of the GRAPE hardware the speed advantage of the
tree-codes compared to direct $N$-body methods was significantly
reduced. The tree-code method speeds up the gravity computation, but it
still forms the major part of the total computation time.  
It is therefore beneficial to execute these
computations using the GRAPE hardware.  By modifying the tree-walk and
using a special version of the GRAPE chip, Fukushige et
al.~\cite{1991PASJ...43..841F} were able to execute the computation of
the gravitational force of the Barnes \& Hut tree-code algorithm using
the GRAPE hardware, thereby benefiting from both the fast tree-code
algorithm and the efficiency of the GRAPE.\\

The first result of a tree-code accelerated by a GPU was presented
in~\cite{2008NewA...13..103B}. The results did show a speed-up compared
to the CPU results, however the speed-up was smaller than with direct
$N$-body methods.  This is of course understandable
since there are fewer force computations that can benefit from the GPU
compared to direct $N$-body methods. Another limiting factor is the
amount of communication required between the GPU and CPU in the
standard GRAPE tree-code implementations. The high computational speed
of the GPU means that communication over the PCI bus can become a
bottleneck. In their award-winning papers Hamada et
al. \cite{citeulike:4604010, Hamada:2010:TAN:1884643.1884644} reduced
this overhead by combining multiple tree-walks (executed on the CPU) 
and transferred these to the GPU in one data transfer. Instead of receiving one set of
interaction lists the GPU now receives more than one which increases
the amount of parallel work that can be executed and improves the
overall efficiency of the GPU. However even with this method the
tree-walk is executed on the CPU.
In {\tt Octgrav}~\cite{OctGravICCS10}, the authors execute the tree-walk 
on the GPU, thereby removing the need to transfer interaction lists 
between the host and the GPU. Furthermore, since this is a data intensive 
operation it benefits from the high on device bandwidth of the 
GPU, which is an order of magnitude higher than that of the CPU.

By removing the traditional bottlenecks of the tree-code algorithm the
performance of the code will become limited by new bottlenecks. Parts
of the algorithm that took only a few percent of the execution time in
the original algorithm suddenly take up a major part of the execution
time. In the tree-code method, for example, the
construction of the hierarchical data-structure, particle
sorting~\cite{169640} or even just the prediction of the new particle
positions become the new bottlenecks. There is only so much one can do
to optimize and resolve bottlenecks
in an algorithm that scales as ${\cal O}(N)$ . In the hierarchical GPU
tree-code {\tt Bonsai} \cite{2012JCoPh.231.2825B} the authors
identified the bottlenecks in {\tt Octgrav} and implemented these on
the GPU. This way the algorithms retained their ${\cal O}(N)$ and ${\cal O}(N \log N)$
scaling,
but profit from the high computational speed and bandwidth of the
GPU. To prevent any further limits they took it one step further and
implemented all parts of the tree-code algorithm on the GPU. This 
eliminated the need to transfer large amounts of data between the CPU
and GPU during each time-step. Using this method the authors are able
to develop a GPU tree-code that works efficiently with a shared time-step
but also in block time-step hierarchy.

In the parallel version of {\tt Bonsai} the authors need the CPU in
order to communicate with other nodes that are connected with each
other in a non-shared memory architecture. By overlapping the
(network) communication time with computations on the GPU the authors
are able to hide most of the required communication overhead of the
relatively slow PCI-bus and network cables.

\section{BRIDGE; Combining Tree and direct}

The last few years have seen a huge increase in computational power
in the form of special purpose hardware and new supercomputers.
And the difference between particle numbers used in collisional and
collisionless methods is only increased.  Depending on the problem
scientists either choose for high precision direct $N$-body methods or
for large particle numbers using approximation methods like the
tree-code. Recently, however, methods have been introduced that try to
combine the best of both worlds. The high accuracy of direct methods
and the speed of tree-codes. In the {\tt BRIDGE} algorithm Fujii et
al.~\cite{2007PASJ...59.1095F} combine a direct $N$-body method and a
tree-code to integrate the evolution of star clusters (which requires direct
$N$-body methods) embedded in their host galaxy (which requires an
approximation method because of the large number of particles). This
allows for detailed simulations in the area of interest while still 
being able to use large
particle numbers. Since the method is based on two well known
algorithms it is possible to use existing tools to speed-up the
method using GPUs.

\begin{figure*}
 \includegraphics[width=\columnwidth]{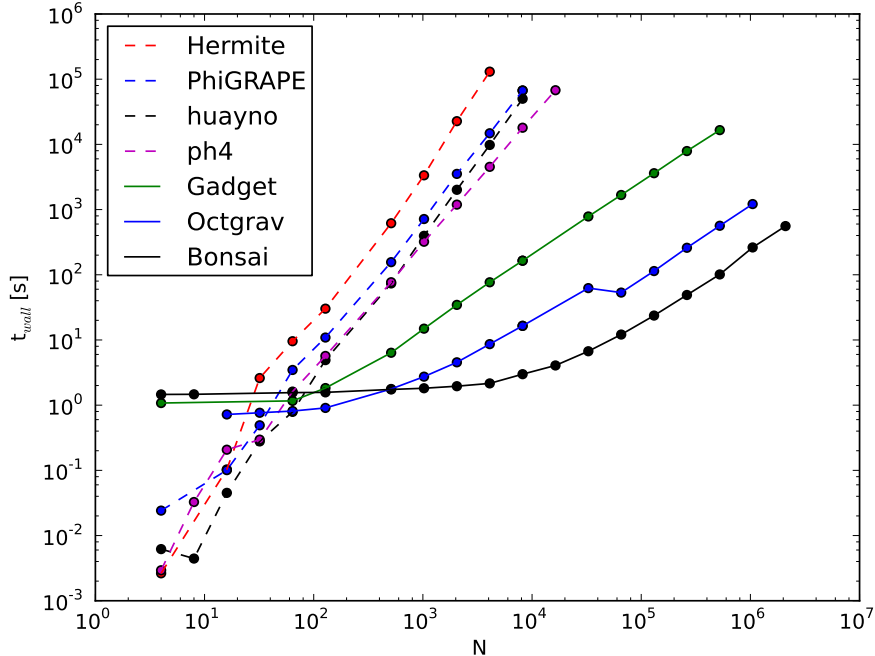}
 \caption{Performance comparison of a suite of $N$-body codes. (Taken from S. PZ 2012 in prep.)}
 \label{fig:NbodyPerformance}        
\end{figure*}

With simulation codes becoming more complex and having more advanced features
it becomes difficult to add new physics to existing codes without breaking 
other parts of the codes. This is a common problem in computational sciences and 
software development in general. Ideas often start off simple, but when something
works you want to extend it, which complicates matters. In AMUSE~\cite{2009NewA...14..369P} a different 
approach is taken. Codes that are written for different specific purposes are 
combined into one framework. This simplifies the development of the separate
software products. The other advantage is that you can combine simulation codes
that have support for GPUs with codes that do not and thereby still 
have the speed advantage of using GPUs. 
With AMUSE it is possible to use the same script using different simulation codes
and thereby having the choice between speed, accuracy or available 
hardware. An example of this is shown in Fig.~\ref{fig:NbodyPerformance} where
the execution speed of a set of $N$-body integration codes is demonstrated.
The figure shows the results of 4 direct $N$-body codes ({\tt Hermite, PhiGRAPE,
Huayno and ph4})  and 3 tree-codes ({\tt Gadget2, Octgrav, Bonsai}). Clearly
visible is the difference in speed and scaling between the direct codes(${\cal O}(N^2)$ scaling)
and the tree-codes (${\cal O}(N \log N)$ scaling).

\section{The future}

With focus shifting to more complex methods and algorithms we see the
advantage of the versatility of GPUs and the shift from fixed function
methods in the early 90s (like the GRAPE) to programmable chips like GPUs.
Even though FPGAs have been around for decades their programming is
difficult and expensive, certainly compared to chips that are
programmable by software. It is much easier to develop and acquire
chips like GPUs, since you can buy them in the computer store around
the corner. The availability and price makes the GPU one of
the most attractive high performance computing devices that are currently
available. It is of course still possible to develop faster chips that
require less energy if you make them dedicated, but the development
cost and specialized knowledge to build a chip that is competitive
against the multi-billion dollar gaming industry is higher than a
university research team can afford.

Also, simulation algorithms become more advanced and incorporate different
techniques to overcome the painful ${\cal O}(N^2)$ scaling. An example
of this is the {\tt Pikachu} code by  Iwasawa et al. (in prep.).
In the {\tt BRIDGE} method one has to indicate which particles 
will be integrated using the direct algorithm and which particles 
with the tree-code algorithm when the initial conditions are created.
The {\tt Pikachu} code improves on this by dynamically deciding which particles
can be integrated using a tree-code and which need direct $N$-body 
integration.

Even though approximation methods (tree-codes, FMM and Particle Mesh) are much
faster than direct $N$-body methods they do not reach the same level
of accuracy. With the increase in computational power, direct $N$-body
methods will always be used for new simulations with increasing $N$ either to
compare to previous results (e.g. performed with approximate methods) or for
new science. The same is valid for the methods used to improve the
performance of direct $N$-body simulations (block time-steps,
neighbour schemes, etc.; see Section~\ref{Sect:1960-1986}).
These all have an influence on the precision.
Although the difference is smaller than the difference between direct
methods and approximations methods it still might be of influence,
especially considering the chaotic nature of the $N$-body
problem~\cite{1964ApJ...140..250M,1993ApJ...415..715G}. 
Therefore, with the increased compute performance we will not
only perform simulations with larger $N$, but also much more detailed
simulations with relatively small $N$ to validate previously obtained
results. Simulations of globular clusters using high precision 
shared time-step algorithms are still far out of reach, but one 
day we will have the computational power to perform exactly 
this kind of simulations.

\begin{figure*}
 \includegraphics[width=\columnwidth]{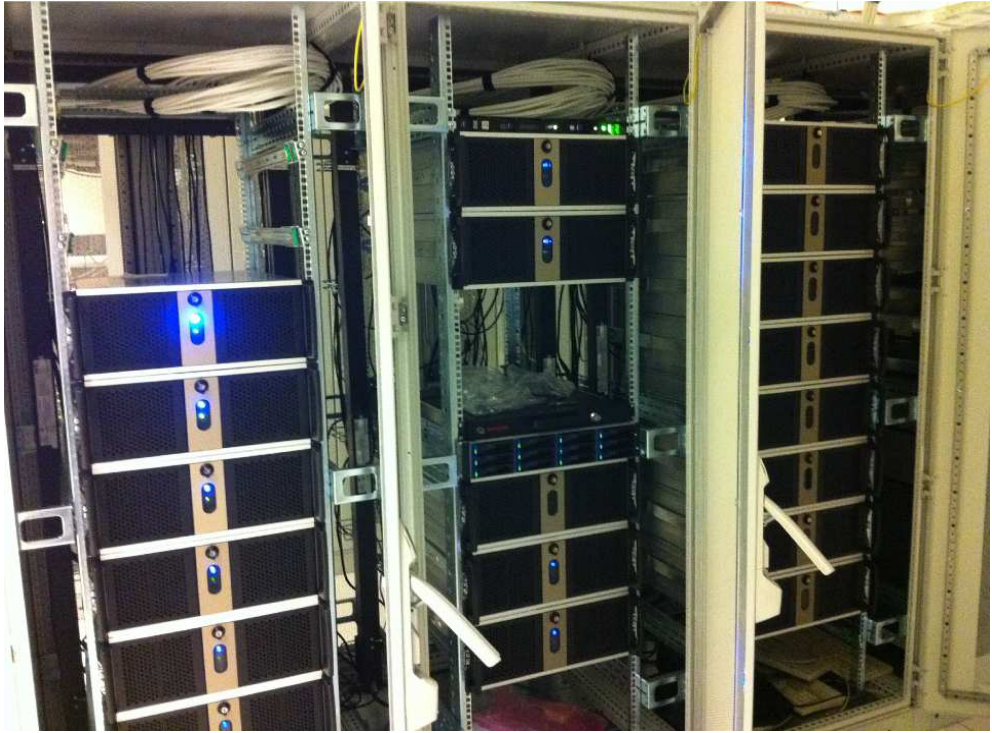}
 \caption{The Little Green Machine. The Leiden GPU cluster.}
 \label{fig:LGMGPUCluster}        
\end{figure*}

The increasing availability of GPUs in supercomputers and in small
dedicated GPU clusters (Fig.~\ref{fig:LGMGPUCluster}) shows the 
potential, increased usage and the faith of researchers in  
GPUs over the last few years. And especially with the installation
of GPUs in ordinary desktop computers, as is done, for example, at the 
Leiden Observatory, this computational power is available at everyone's
fingertips without having to request time on expensive supercomputers.

% \section*{Acknowledgements}

\begin{acknowledgement}
This work was
supported by NWO (grants \#643.000.802, VICI [\#639.073.803], AMUSE [\#614.061.608]
and LGM [\# 612.071.503]), NOVA and the LKBF.
\end{acknowledgement}

% \bibliographystyle{plainnat}
% \bibliography{intro}

% Het reference moet waarschijnlijk als volgt : 
% \begin{thebibliography}{}
% Author, Journal \textbf{Volume}, (year) page numbers
% Author, \textit{Book title} (Publisher, place year) page numbers
% etc
% \end{thebibliography}

\end{document}